\documentclass{easychair}

\usepackage{doc}
\usepackage{multirow}
\usepackage{makecell}
\usepackage{float}
\usepackage{url}
\usepackage{comment}

\title{How many bots are you following?%
\thanks{Partially supported by the European Union’s Horizon 2020 programme (grant agreement No. 830892, SPARTA) and by IMT Scuola Alti Studi Lucca: Integrated Activity Project TOFFEe ‘TOols for Fighting FakEs’.}
}

\author{
Alessandro Balestrucci\inst{1,2}
}

\institute{
IMT School for Advanced Study Lucca, Lucca, Italy\\
\and
 Gran Sasso Science Institute,
 L'Aquila, Italy\\
 \email{alessandro.balestrucci@gssi.it}
 }

\authorrunning{A. Balestrucci}

\titlerunning{How many bots are you following?}

\begin{document}

\maketitle
\begin{abstract}
Social Media are evolving as a pervasive source of news
able to reach a larger audience through their spreading power. The main drawback is given by the presence of malicious accounts, known as social bots, which are often used to diffuse misleading information. Social bots are automated accounts whose goal is to interact with humans and influence them.
Starting from the definition of \textit{credulous} (i.e., human accounts with a high percentage of bot friends among their \textit{followees}), in this work we aim to single out a regression model to derive, with an acceptable margin of error, the percentage of bot-followees of a human-operated account. 
The advantage lies in knowing, as a preventive measure, which users may be the target of bots' activities, hence more exposed to the misleading/unreliable content. Our results showed that
the best regression model achieves a Mean Absolute Error of 3.62\% and a Root Mean Squared Error of 5.96\%, thus encouraging further research in this direction. 
\\\\
\textbf{keywords} -- Credulous Twitter Users, Humans-Bots Interactions, Disinformation Spreading,  Social Networks Analysis, Supervised Learning 
\end{abstract}

\section{Introduction}
\label{sec:intro}
The widespread use of mobile devices, combined with the ease and cheapness of being connected, are some of the most important factors increasing the usage of Online Social Media (OSM) as means of communication~\cite{newman2016digital}. Several studies~\cite{costera2015checking,newman2017reuters,aneez2019reuters} recently confirmed OSM (such as Facebook and Twitter) as more pervasive than traditional mass media (e.g., radio, newspaper, etc.), especially among youngest people~\cite{newman2016digital}. 

As consequence of the OSM's pervasiveness, a larger audience is reached, but several issues arise about the veracity of the circulating news.
This leads to fake news and misinformation that are intentionally propagated for harmful purposes~\cite{lazer2018science}. 
The information diffusion on OSM is often supported by automated accounts controlled, totally (bots) or partially (cyborg), by ad-hoc software applications~\cite{Varol17}. Under fictive identity, such automated accounts actively interact (social bots~\cite{ferrara2016rise}) with genuine users and share/produce contents of doubtful credibility. 
Governments\footnote{\url{https://tinyurl.com/yym2xa3v}}, academics~\cite{zhou2018fake} and OSM administrators\footnote{\url{https://tinyurl.com/yac7lsn6}} are struggling to face these problems.

Despite the efforts spent by OSM administrators in removing suspicious accounts\footnote{Facebook: \url{https://tinyurl.com/y3yzvpah}}\footnote{Twitter: \url{https://tinyurl.com/y3efs8s5}}, and by researchers in improving bot detection techniques~\cite{mushtaq2019distributed,amina2019bibliometric}, this plague is far from being eradicated. In fact, it has been estimated that, on Twitter, the percentage of bots among the active users ranges between 9\% and 15\%~\cite{Varol17}. Furthermore, a new generation of bots, more sophisticated than the previous ones, are able to avoid detection by mimicking human behaviour~\cite{daniel2019bots}.\\
In some cases, bots have been used to call volunteers in case of emergencies~\cite{savage2016botivist} or to spread academic events such as conferences, but these are only exceptions. In fact, a dominant use of these artificial entities is for malicious purposes, e.g., to encourage hate speeches, misconception and, more in general to influence people~\cite{suarez2016influence,keller2019social}.\\
The effectiveness of such malicious activities, intended to manipulate public opinion, has been investigated on the Brexit referendum~\cite{howard2016bots}, the US Presidential election 2016~\cite{bessi2016social}, the elections in France~\cite{ferrara2017disinformation} and Mexico~\cite{bradshaw2017troops}. These are few well-known examples of how and to what extent bots and disinformation can damage democracy, especially if this kind of campaigns target specific categories of users.\\
A recent study stated that the majority of genuine (human) users usually do not check the reliability of contents on OSM, and many even share these articles~\cite{efthimion2018supervised}; hence contributing, although unknowingly, to diffuse unreliable content. Moreover, in~\cite{kempe2003maximizing} models for influence propagation in OSM have been studied, and a strong correlation emerged between the target nodes (to influence) and the role of their neighbours (social contacts).

Using Twitter as benchmark, this paper focuses on finding a way to determine, as precisely as possible, the amount of potential malicious accounts (bots) a human is following (\textit{bot-followees}).
In our opinion, the more bots a user is following, the more she/he is exposed to potentially malicious activities.
Specifically, by using Machine Learning (ML) techniques, we aim to single out a regression model able to predict, for a human-operated account, the percentage of its \textit{bot-followees}. 
Starting from a publicly available dataset\footnote{Dataset: \url{https://tinyurl.com/y4o98c7l}} of \textit{credulous} users, built in~\cite{balestrucci2019you}, and by using commonly employed features from the literature~\cite{Varol17,Cresci15fame}, we build several regression models by means of various ML algorithms. 
We then use Mean Absolute Error (MAE) and the Root Mean Squared Error (RMSE) for evaluation. 

For the sake of completeness, we experiment on two ``versions'' of the considered dataset: firstly by considering the subset of credulous users only (with 316 instances, henceforth called \textit{credulous-only}), and then on the whole set of human-operated accounts (with 2,838 instances, named \textit{all\_humans}. 
The experimental results are promising. The best model trained by using \textit{credulous-only}, achieves a MAE of 4.32\% and a RMSE of 6.62\%. On the other hand, astonishingly, the best model trained on the whole humans set, fits better; in fact it obtains a MAE of 3.62\% and a RMSE of 5.96\%. Although RMSE scores are not the best obtained, by following the finding  in~\cite{willmott2005advantages}, we prefer to give more importance to MAE scores than to RMSE.

Despite these encouraging results, additional efforts are needed to improve users' awareness. 
The usefulness in improving the accuracy of these models is twofold. Firstly, in a preventive way, to identify sensible users more exposed to the malicious activities of bots and by being identified as good targets of misinformation campaigns. Secondly, to safeguard the usefulness, credibility and effectiveness of social media as mean of communication.

The remainder of this paper is organized as follow: Section \ref{sec:rw} presents the related work; Section \ref{sec:expset} exposes the general approach and provides the details of the performed experiments;
Section \ref{sec:expres} illustrates the experimental results; Section \ref{sec:disc} discusses the main findings, and Section \ref{sec:conc} concludes the paper by outlining future research.

\section{Related Work}
\label{sec:rw}

This paper aims to investigate the dynamics and causes of human users influenced by Social Media, especially if interacting with bots and exposed to misinformation. In the following, we review the related work dealing with similar goals.

Despite this cannot be considered an exhaustive literature review, we tried to supply concepts, definitions and adopted strategies related to interactions between human-operated accounts and bots on Social Media.

In~\cite{wagner2012social} the authors stress the importance of adopting protection mechanisms in Social Media to protect human-operated accounts. To this purpose, the authors provided the definition of \textit{susceptible} users, i.e., users who start to interact with a social bot (even if only once). By introducing a set of 97 features (categorized in \textit{linguistic}, \textit{network} and \textit{behavioural}), a binary classifier is successfully produced to spot out \textit{susceptible} users. A regression model was also built to predict users level of \textit{susceptibility} without success. 
Inspired by the definition of \textit{susceptible} users and finding the regression model construction a very challenging goal, in our work we investigate the usage of a regression model able to predict the percentage of bots followed by a human-operated account.

On the same line of the described research, in~\cite{wald2013predicting} the authors conducted a feature study to single out those features useful to predict whether a user is likely to interact with a bot. To this end, the considered users have been contacted by a bot through a tweet mention, and when replying users are labelled as susceptible. The drawback of this approach is represented by the key feature (i.e., \textit{klout} score), no longer available due to the related web-service closure.

In~\cite{mitter2014categorization} a comprehensive categorization scheme for social bot attacks in Twitter has been proposed. The analysis of each attack is considered along different dimensions, i.e., \textit{targets, account types, vulnerabilities, attack methods} and \textit{results}. This categorization is useful to investigate to which class of attacks users with a considerable percentage of \textit{bot-followees} are vulnerable. 
Specifically, the authors observed the impact of social bots in link creation between targeted human-operated accounts in Twitter. 

In~\cite{DBLP:conf/websci/ShenCGLYL19}, inspired by~\cite{wagner2012social}, \textit{gullible} users have been defined in relation to their susceptibility to fake news. 
The authors present five degrees of susceptibility (referred to a user's reply to a fake news) and, by using the same features of~\cite{wagner2012social}, attempt to build a multi-class classifier predicting the user's reply susceptibility level. The classifier achieves an AUC of 0.82. Similarly to this approach, we also used existing feature sets (Botometer~\cite{davis2016botornot} and ClassA~\cite{Cresci15fame}), but applied to a regression task.

An interesting case study is presented in~\cite{sandy2017can}, where human's skills have been tested in recognizing fake accounts generated by the M3 app~\cite{li2014hiding}. In general, there is a good humans' ability to distinguish the generated accounts, but there are no details on the characteristics of such fake accounts. 
Sharing the same motivation, our work is related to the production of a regression model that, complementary to~\cite{sandy2017can}, quantifies when a human user is not able to recognize a bot.

We conclude this section by mentioning our previous works~\cite{balestrucci2019identification,balestrucci2019you}.
Beside the definitions of \textit{susceptible} users~\cite{wagner2012social} and \textit{gullible} users~\cite{DBLP:conf/websci/ShenCGLYL19}, in~\cite{balestrucci2019identification} we introduced the concept of \textit{credulous} users, i.e., human operated accounts following a considerable amount of bots (i.e., \textit{bot-followees} using Twitter's slang). 
A set of rules have been introduced to discern whether a user is a credulous, and these rules  allow to rank human operated accounts by relying on the ratio of bots (supported by a bot detector) over their followees. On a dataset with more 700 human operated accounts, only 64 have been identified as credulous\footnote{First Credulous Dataset: \url{https://tinyurl.com/y6lod2yz}}. Here, the drawback is represented by the huge amount of data needed to investigate the followees of humans under analysis.

To overcome this issue, in~\cite{balestrucci2019you}, we built a binary classifier to find out credulous Twitter users by considering a larger \textit{ground-truth} (of almost 3k\footnote{Dataset: \url{https://tinyurl.com/y4o98c7l}}) and considering only features of the profile and not of users' followees. 
We ended up with a lightweight (in terms of costs for gathering the data) classifier. The achieved classification performance have been very promising with an accuracy of 93.27\% and an \textit{AUC} (Area Under the ROC curve) of 0.93.

Although this work is in line with the themes of the previous ones~\cite{balestrucci2019identification,balestrucci2019you}, it differs in scope and strategy. The primary target is the prediction of a percentage value of bot-followees, and the adopted strategy is no longer the classification, but the regression. 

\section{Experimental Setup}
\label{sec:expset}

This section explains the dataset, the features and the experimental design; specifying the metrics used to evaluate the performance of the trained models.

\subsection{Dataset}
The considered dataset includes 2,838 IDs of human-operated accounts on Twitter, presented in~\cite{balestrucci2019you} and publicly available\footnote{Dataset: \url{https://tinyurl.com/y4o98c7l}}. A label (0 or 1) is associated to each entry   to indicate whether an account is \textit{credulous} (with 1) or not (with 0).\\ 
For each account we downloaded all the information (timeline, tweet mentions and profile data) of its \textit{followees}. Then, through of a bot detector~\cite{balestrucci2019you}, it has been possible to assign, to each \textit{followee}, a label indicating whether it is a bot. This allows us to derive a ground-truth where each accounts is assigned to a percentage of its bot-followees. As anticipated in Section \ref{sec:intro}, two versions have been built: one including the 316 entries labeled as credulous (\textit{credulous-only}) and another one including all 2,838 human-operated accounts (\textit{all\_hums}).

All accounts in our dataset belong to human users and are taken from the three publicly available sources\footnote{BotRepository:\url{https://tinyurl.com/yxfxmqac}} briefly described below:
\paragraph{cresci-2015~\cite{Cresci15fame}.} 
This repository consists of three smaller datasets: (i) 469 Twitter accounts certified as human-operated during in a research project named  \textit{@TheFakeProject}. 
%
(ii) 1,481 accounts selected as genuine users as the result of a sociological study relying on manual verification.
(iii) 833 fake accounts, bought from three different Twitter accounts online markets.
%
\paragraph{cresci-2017~\cite{cresci17paradigm}.} This repository contains 3,474 Twitter accounts certified as humans and 6,609 social spambots (e.g., spammers of job offers and advertising products on sale at Amazon). The humans were selected through a hybrid crowd-sensing approach~\cite{avvenuti2017}; the authors of~\cite{cresci17paradigm} randomly contacted Twitter users and asked simple questions in natural language, all replies were then manually verified.

%
%
\paragraph{varol-2017~\cite{Varol17}.} 
This repository consists of 2,573 Twitter accounts  
selected after a manual annotation based on inspecting the profile details and the produced content. Overall, 1,747 Twitter accounts were annotated as human-operated and 826 as bots.\\


\subsection{Features}
\label{subsec:feat}
The features employed to represent the human accounts can be grouped in three sets. The first set inherits the features from the Botometer web service\footnote{Botometer web service (RapidAPI): \url{https://tinyurl.com/yytf282s}}, plus some additional ones introduced in~\cite{balestrucci2019identification}. We use the Botometer \textit{macro categories scores}, which encompasses the essence of all the features (around 1,000)~\cite{davis2016botornot}, on which the outputs' scores are computed. We extend this set by considering also: the CAP (Completely Automated Profile) value, 
the ``Scores''~\cite{yang2019arming}, the number of tweets and mentions. This augmented set of features will be called \textit{Botometer+} in the rest of the paper. 

The second set inherits the so-called \textit{ClassA} features, singled out in~\cite{Cresci15fame}. We discard the one considering \textit{duplicated pictures}, because we could not verify whether the picture of a profile was also used in another one. 
This subset is called \textit{ClassA-}. We want to stress that these features can be extracted by only looking at the social profile of an account, without considering the timeline, that is instead considered in \textit{Botometer+}.
A schematic representation of these two feature sets is reported in table~\ref{tab:features} in  Appendix~\ref{app:feat}.

The third set of features has been built through the union of the two previous ones and called \textit{ALL\_features}. We want to remark that these three feature sets have been successfully used for binary classification task in~\cite{balestrucci2019you}; and the introduction of new features is out of the scope of this work. 

\subsection{Experimental Design}
\label{subsec:expdes}
The experimental session starts by setting up the data; precisely, by transforming the entries of our dataset accordingly to the three sets of features. 

Afterwards, to train regression models, 14 algorithms have been employed. The machine learning framework utilized to conduct the experiments is Weka~\cite{hall2009weka}.
Accordingly to the notation adopted by the tool, the algorithms are: \texttt{ZeroR}\footnote{ZeroR weka: \url{https://tinyurl.com/y4hdhp54}} (used to obtain a baseline value against which to compare the values of the other model), REPTree~\cite{quinlan1987simplifying}, LinearRegression~\cite{le1992ridge}, k-Nearest Neighbour (IBk)~\cite{aha1991instance}, LWL~\cite{Atkeson1996}, AdditiveRegression~\cite{Friedman1999}, RegressionByDiscretization~\cite{frank2009conditional}, M5Rules~\cite{Quinlan1992}, DecisionStump~\cite{iba1992induction}, GaussianProcess~\cite{Mackay1998}, SMOreg~\cite{Shevade1999}, MultilayerPerceptron~\cite{pal1992multilayer}, MLPRegressor\footnote{MLPRegressor weka: \url{https://tinyurl.com/y5krc6d2}}, RandomForest~\cite{breiman2001random}.
To investigate whether regression models, built on \textit{credulous} users work better than the general case, we experiment on both versions (\textit{credulous-only} and \textit{all\_hums}).

\paragraph{Results evaluation}\label{par:resEval} 
At best of our knowledge, there are few papers addressing this problem~\cite{wagner2012social,DBLP:conf/websci/ShenCGLYL19}. Hence, no well-defined baseline is available from the literature to compare our results. To overcome this issue, we compare the performance of the trained models (obtained from the aforementioned algorithms) with the score calculated by \texttt{ZeroR} method, since it predicts, for each human-operated account, the average value of the population.\\
To evaluate models' performance, two metrics have been used: Mean Absolute Error (MAE) and Root Mean Squared Error (RMSE) that are widely used for regression tasks\footnote{\url{https://tinyurl.com/yd9ljcmj}}~\cite{moolayil2019learn}. The former measures the average of errors in a set of numerical predictions, between the real values and the predicted ones ($ MAE =\frac{1}{n}\sum_{j=1}^{n}{|y_{real_j} - y_{pred_j}|}$); the latter measures the error too, but it stresses more the prediction error by raising the square of the difference between the real values and the predicted ones ($ RMSE =\sqrt{\frac{1}{n}\sum_{j=1}^{n}{(y_{real_j} - y_{pred_j})^2}}$). 

Through the experimental results analyzer embedded in Weka~\cite{scuse2007weka}, 
we performed statistical tests (paired T-Test~\cite{hsu2007paired} with $\alpha=0.05$) to determine which algorithm performs significantly better than the baseline, relatively to each set of features. 

All the results are presented in the following section~\ref{sec:expres} and then discussed in section~\ref{sec:disc}.
\section{Experimental Results}
\label{sec:expres}

For sake of clarity, 
Section~\ref{subsec:cred} reports the results referring to the experiments performed on the set of credulous users only (\textit{credulous-only}). 
Section~\ref{subsec:HOA} shows the outcome derived by considering all the human-operated accounts (\textit{all\_hums}).

Tables~\ref{tab:rmse_cred}-\ref{tab:mae_allHum} have the same structure, differing in evaluation metric and ground-truth data entries.
In the first column the algorithms mentioned in Section~\ref{subsec:expdes} are listed. 
The remaining columns show the scores obtained when dataset's instances are represented according to a specific feature set (see section~\ref{subsec:feat}) namely \textit{Botometer+}, \textit{ClassA-} and \textit{All\_features}. The first row of the table contains the baseline obtained through the \textit{ZeroR} method.
%
The star symbol indicates some tables' entries whose associated value is significantly lower than the baseline, according to the paired t-test performed by Weka (see results evaluation in Section~\ref{subsec:expdes}). The lowest score is reported in bold.


\subsection{Credulous-only}
\label{subsec:cred}
Table~\ref{tab:rmse_cred} reports the scores related to the RMSE. The baseline is offered by ZeroR showing a score equal to 6.73\%.
%
Note that in the column headed \textit{Botometer+}, there are no starred values. Despite some of them are lower than the baseline (i.e., \textit{LinearRegression}, \textit{M5Rules}, \textit{GaussianProcesses}, \textit{SMOreg} and \textit{RandomForest}), their values have not been considered significantly lower by Weka, in fact they are slightly better than the baseline. Among all values, the lowest one belongs to \textit{GaussianProcesses} (6.48\%) followed by \textit{LinearRegression} (6.52\%). 
\begin{table}[htbp]
	\begin{centering}
		\begin{tabular}{lccc}
		\hline
		\multirow{2}{*}{\textbf{Algorithms}} & \multicolumn{3}{c}{\textbf{Feature sets}}\\
		\cline{2-4}
		& \textit{Botometer}+ & \textit{ClassA-} & \textit{All\_features} \\
		\hline
		$\rightarrow$ ZeroR $\leftarrow$ (baseline) & 6.73 & 6.73 & 6.73 \\
        REPTree & 6.92 & 6.86  & 6.86  \\
        LinearRegression & 6.52 & 8.52  & 8.62  \\
        IBk & 8.71 & 8.95  & 8.02  \\
        LWL & 6.84 & ~\textbf{6.10}* & 6.26  \\
        AdditiveRegression & 6.79 & 6.30  & 6.20  \\
        RegressionByDiscretization & 8.43 & 7.47  & 7.78  \\
        M5Rules & 6.53 & 9.20  & 7.44  \\
        DecisionStump & 6.90 & ~6.15* & ~6.15* \\
        GaussianProcesses & 6.48 & 7.34  & 7.55  \\
        SMOreg & 6.62 & 7.70  & 7.71  \\
        MultilayerPerceptron & 10.79 & 11.97  & 11.82  \\
        MLPRegressor & 7.59 & 7.50  & 6.86  \\
        RandomForest & 6.60 & ~6.15* & 6.21  \\
		\hline
		\end{tabular}
		\caption{Root Mean Squared Error (RMSE) on credulous users}
		\label{tab:rmse_cred}
	\end{centering}
\end{table}


Differently from the \textit{Botometer+} case, the column \textit{ClassA-} there are some starred values. 
Both \textit{DecisionStump} and \textit{RandomForest} achieve a RMSE score of 6.15\%; but, the lowest value of 6.10\% belongs to \textit{LWL}, which also is the best score reported in Table~\ref{tab:rmse_cred}.

The last column of Table \ref{tab:rmse_cred}
contains only one starred value which also the column's lowest score (\textit{DecisionStump}). Like for the previous column, also here there are some values lower than the baseline (\textit{AdditiveRegression} with 6.20\% and \textit{RandomForest} with 6.21\%), but not enough to be labeled as star entries.

\vspace{1em}
Table~\ref{tab:mae_cred} exposes the scores related to MAE. The indicated baseline is of 4.84\%. 

Looking at \textit{Botometer+} features, the values lower than the baseline are: 4.83\% (\textit{LWL}, \textit{AdditiveRegression}, \textit{DecisionStump}), 4.68\% (\textit{LinearRegression}, \textit{M5Rules}), 4.66\% (\textit{GaussianProcesses}) and 4.32\% (\textit{SMOreg}, starred). The latter is the lowest MAE.

When considering \textit{ClassA-} features, the values lower than the baseline are: 4.78\% (\textit{LinearRegression}), 4.64\% (\textit{SMOreg}), 4.55\% (\textit{AdditiveRegression}), 4.54\% (\textit{RandomForest}) and 4.36\% (\textit{DecisionStump} and \textit{LWL}, both starred). 

If all features are analyzed, the values lower than the baseline are: 4.78\% (\textit{MLPRegresso}), 4.77\% (\textit{REPTree}), 4.67\% (\textit{SMOreg}), 4.44\% (\textit{RandomForest}), 4.40\% (\textit{LWL}, starred), 4.39\% (\textit{AdditiveRegression}, starred), 4.36\% (\textit{DecisionStump}, starred).

\begin{table}[tbp]
	\begin{centering}
		\begin{tabular}{lccc}
		\hline
		\multirow{2}{*}{\textbf{Algorithms}} & \multicolumn{3}{c}{\textbf{Feature sets}}\\
		\cline{2-4}
		& \textit{Botometer}+ & \textit{ClassA-} & \textit{All\_features} \\
		\hline
		$\rightarrow$ ZeroR $\leftarrow$ (baseline) & 4.84 &  4.84 & 4.84 \\
        REPTree & 4.91 &4.78   & 4.77   \\
        LinearRegression & 4.68 &5.18   & 5.19   \\
        IBk & 5.88 &6.17   & 5.44   \\
        LWL & 4.83 & ~4.36*& ~4.40*\\
        AdditiveRegression & 4.83 &4.55   & ~4.39*   \\
        RegressionByDiscretization & 5.88 &5.22   & 5.54   \\
        M5Rules & 4.68 &5.18   & 4.90   \\
        DecisionStump & 4.83 & ~4.36* & ~4.36* \\
        GaussianProcesses & 4.66 &4.90   & 4.95  \\
        SMOreg & ~\textbf{4.32}* &4.64 & 4.67  \\
        MultilayerPerceptron & 6.91 &8.17   & 7.92  \\
        MLPRegressor & 5.16 &5.19   & 4.78   \\
        RandomForest & 4.86 &4.54   & 4.44   \\
		\hline
		\end{tabular}
		\caption{Mean Absolute Error (MAE) on credulous users}
		\label{tab:mae_cred}
	\end{centering}
\end{table}

\subsection{All humans}
\label{subsec:HOA}
In this Section we report
the results related to the experiments performed on all human-operated accounts (2,838 instances).

Table~\ref{tab:rmse_allHum} reports the values concerning RMSE, with a baseline value of 6.25\%.

\begin{table}[htbp]
	\begin{centering}
		\begin{tabular}{lccc}
		\hline
		\multirow{2}{*}{\textbf{Algorithms}} & \multicolumn{3}{c}{\textbf{Feature sets}}\\
		\cline{2-4}
		& \textit{Botometer}+ & \textit{ClassA-} & \textit{All\_features} \\
		\hline
		$\rightarrow$ ZeroR $\leftarrow$ (baseline) & 6.25~ & 6.25 & 6.25 \\
        REPTree & ~5.96* & ~6.02* & ~5.93* \\
        LinearRegression & ~5.77* & 6.14 & ~5.80* \\
        IBk & 7.73  & 8.58  & 7.59  \\
        LWL & ~5.91* & 6.08 & ~5.99* \\
        AdditiveRegression & ~5.84* & 6.07  & ~5.80* \\
        RegressionByDiscretization & 6.32  & 6.43  & 6.83  \\
        M5Rules & ~6.02*  & 6.16  & ~5.84* \\
        DecisionStump & ~5.96* & ~6.06* & ~6.02*\\
        GaussianProcesses & ~5.79* & 6.13 & ~5.83* \\
        SMOreg & ~5.91* & 6.36  & ~5.96* \\
        MultilayerPerceptron & 7.67  & 6.53  & 9.47  \\
        MLPRegressor & ~5.89* & ~6.06*  & 6.84  \\
        RandomForest & ~5.92* & 6.09  & ~\textbf{5.72}* \\
		\hline
		\end{tabular}
		\caption{Root Mean Squared Error (RMSE) on all the human-operated accounts}
		\label{tab:rmse_allHum}
	\end{centering}
\end{table}

In the second column (\textit{Botometer+}), almost all the RMSE scores are lower than the baseline and starred, with the exception of \textit{IBk} (7.73\%), \textit{RegressionByDiscretization} (6.32\%) and \textit{MultilayerPerceptron} (7.67\%). With a score of 5.77\%, \textit{LinearRegression} has the lowest column's RMSE (and the second better one in Table~\ref{tab:rmse_allHum}).
Concerning the third column (\textit{ClassA-}), the situation is slightly worse. Despite nine values have better scores than the baseline, only three of them are significantly lower: 6.02\% (\textit{REPTree}, the lowest) and 6.06\% (\textit{DecisionStump} and \textit{MLPRergessor} both). 
The forth column (\textit{ClassA-}) shows a situation close to \textit{Botometer+}'s case. In fact, with exception of four cases (i.e., \textit{IBk}, \textit{RegressionByDiscretization}, \textit{MultilayerPerceptron} and \textit{MLPRegressor}), all the other entries have significantly better values w.r.t. baseline (starred table's entries). The lowest RMSE of 5.72\% is achieved by using \textit{RandomForest} and it is the better score of Table~\ref{tab:rmse_allHum}.

\vspace{1em}
Finally, Table~\ref{tab:mae_allHum} presents the MAE outcome. The calculated baseline is 4.21\%. 
\begin{table}[htbp]
	\begin{centering}
		\begin{tabular}{lccc}
		\hline
		\multirow{2}{*}{\textbf{Algorithms}} & \multicolumn{3}{c}{\textbf{Feature sets}}\\
		\cline{2-4}
		& \textit{Botometer}+ & \textit{ClassA-} & \textit{All\_features} \\
		\hline
		$\rightarrow$ ZeroR $\leftarrow$ (baseline) & 4.21 &  4.21 & 4.21 \\
        REPTree & ~3.95* & ~3.94* & ~3.87* \\
        LinearRegression & ~3.84* & 4.08 & ~3.83* \\
        IBk & 5.07  & 5.43  & 4.95  \\
        LWL & ~3.97* & ~4.00* & ~3.98* \\
        AdditiveRegression & ~3.89* & ~3.93* & ~3.76* \\
        RegressionByDiscretization & 4.16  & 4.24  & 4.36  \\
        M5Rules & ~3.91* & ~3.96* & ~3.82* \\
        DecisionStump & 4.06 & ~3.99* & 4.07\\
        GaussianProcesses & ~3.87* & 4.09 & ~3.87* \\
        SMOreg & ~3.67* & ~3.84*  & ~\textbf{3.62}* \\
        MultilayerPerceptron & 4.90  & 4.39  & 5.14  \\
        MLPRegressor & ~3.88* & ~3.93* & 4.07  \\
        RandomForest & ~3.96* & ~3.96* & ~3.77* \\
		\hline
		\end{tabular}
		\caption{Mean Absolute Error (MAE) on all the human-operated accounts}
		\label{tab:mae_allHum}
	\end{centering}
\end{table}
At first sight, regardless to the feature sets, almost all the entries are lower than baseline; and most of them are starred. When considering \textit{Botometer+} features, the exceptions are: \textit{IBk} (5.07\%) and \textit{MultilayerPerceptron} (4.90\%); \textit{RegressionByDiscretization} (4.16\%) and \textit{DecisionStump} (4.06\%) are not lower enough to gain the star. For \textit{Botometer+}, the lowest MAE is 3.67\% (the second better value) achieved by the model built by mean of \textit{SMOreg}.

Similarly, when analyzing \textit{ClassA-} features, the values higher than the baseline are: 5.43\% (\textit{IBk}), 4.24\% (\textit{RegressionByDiscretization}) and 4.39\% (\textit{MultilayerPerceptron}). The remaining values are significantly lower than the one in the \textit{ZeroR}'s row but 4.08\% (\textit{LinearRegression}) and 4.09\% (\textit{GaussianProcesses}). Even in this case, the lowest score (3.84\%) belongs to \textit{SMOreg}.\\
Like in the previous cases, even when all features are taken into account, almost all values are lower, not only compared to the baseline, but also compared (by row) to the values of the other two feature sets.
The values overcoming the baseline are: 5.14\% (\textit{MultilayerPerceptron}), 4.95\% (\textit{IBk}) and 4.36\% (\textit{RegressionByDiscretization}). Except \textit{MLPRegressor} and \textit{DecisionStump} (both 4.07\%), all the other entries have significantly better MAE values (starred).
Once again, \textit{SMOreg} outperforms other algorithms' scores for all features, with a MAE of 3.62\%; the lowest in Table~\ref{tab:mae_allHum}.
\section{Discussion}
\label{sec:disc}

It is possible to immediately notice that better results can be obtained when the full version of the dataset (\textit{all\_hums}) is considered rather than \textit{credulous-only}, in terms of both quantity (number of models with lower performance than the baseline) and quality (statistical significance of the values). 

However, focusing on baseline values of each metric, the highest ones are observed in the version \textit{credulous-only}.
This is due to a higher distance between the real values (by credulous class attribute) and the average value calculated on them. 
The fact that \textit{all\_hums}'s baselines have lower values indicates a better closeness to the average value of not-credulous instances.

Considering the MAE metric, the model generated by the SMOreg algorithm is the most accurate, regardless of the dataset versions. 
This concordance is limited exclusively to the algorithm, as these performances have been obtained with two different feature sets: 4.32\% by using \textit{Botometer+} for \textit{credulous-only} version and 3.62\% with \textit{All\_features} for the \textit{All\_hums} version.

But, considering the dataset with all instances, and looking the MAE's score obtained by SMOreg (using \textit{Botometer+}'s features), we can see that the value is very similar (3,67\%, the second-best result). This 0,05\% loss (3,62 vs 3,67) can still be overshadowed by the advantage of not having to calculate the ClassA- features (included in \textit{ALL\_features}). Therefore, at least as far as MAE metrics are concerned, a representation in \textit{Botometer+} features combined with the use of the \textit{SMOreg} algorithm can be considered the best choice.

Good performances are also achieved by the algorithms \textit{LWL} and \textit{RandomForest}, producing the models with the best RMSE in \textit{credulous-only} and \textit{all\_hums}, respectively. \textit{DecisionStump} and \textit{AdditiveRegression} produce good models too, and many times even with errors significantly lower than the baseline. As for the models generated through \textit{REPTree}, \textit{LinearRegression}, \textit{M5Rules}, \textit{GaussianProcesses} and \textit{MLPRegressor}, it is possible to notice an inequality depending on the version of the dataset to which they are applied, proving more effective when used on the full version. Unfortunately, the use of \textit{IBk} \textit{RegressionByDiscetization} and \textit{MultilayerPerceptron} has not been profitable, regardless of both the set of used features and the version of the considered dataset.

In some cases, on the same algorithm, the score obtained by using \textit{All\_features} is identical (or very similar) to the one in \textit{ClassA-} or \textit{Botometer+}. These situations occur when the algorithm prefers to consider the components of a certain feature set. Some examples are given by: \textit{DecisionStump} in Tables~\ref{tab:mae_cred} and~\ref{tab:rmse_cred}, \textit{GaussianProcesses} in Table~\ref{tab:mae_allHum} and \textit{REPTree} in  Table~\ref{tab:rmse_cred}. 

Further findings can be provided by studying, for each evaluation metric, to what extent the choice of a feature set (w.r.t. another one) can affect the overall performance. Regardless of the cost to calculate a feature set, the experimental results do not show a great disparity in preferring one feature set over another. Therefore, as far as the "MAE" is concerned, the previous assertion of preferring the \textit{Botometer+} functionalities remains valid.

On the contrary, considering RMSE, the situation is more complicated. The best values come from the use of two distinct feature sets: \textit{ClasseA-} (\textit{credulous-only}) and \textit{All\_features} (\textit{all\_hums}). By using a similar logic adopted for the MAE, it is possible to notice that, the second-best RMSE value for \textit{credulous-only} is identical both for \textit{ClassA-} and  \textit{All\_features}. Therefore, unlike MAE case, if we concern about the RMSE only, we can prefer \textit{All\_features}, making the calculus of \textit{ClassA-}'s features mandatory.

\vspace{-0.5em}
\section{Conclusion}
\label{sec:conc}

The recent literature shows a great deal of concern regarding the effectiveness of malicious bots and in their role to affect several domains (economics, politics, etc.). Not only as regards the spreading of fake news, but especially on their ability to interact with and deceive human users. In fact, the dissemination of low-credibility content is not only performed by social bots but also emphasized by genuine users by sharing it~\cite{efthimion2018supervised}.

In this work, we devote our attention to single out a method that enables us to quantify, as precisely as possible, the percentage of bots followed (\textit{bot-followees}) by human operated accounts in Twitter. 
By mean of machine learning regression techniques, we derived predictive models. The experimentation has been conducted firstly by considering {\it credulous} users only (i.e., humans with high percentage of bots among their friends) and then on all the human-operated accounts in the dataset.

For our experimentation we considered three feature sets, namely: \textit{Botometer+}, \textit{ClassA-} and their union \textit{All\_features}. 
We recall that \textit{Botometer+}'s features are obtained by using the related web-service, and for each user the following input is required: profile data, tweets, mention-tweets. The calculus of \textit{ClassA-}'s features is less expensive because such features are derived by analyzing users' data profile only. 
The best regression model, obtained on the complete dataset trained with the \textit{SMOreg} algorithm, achieves promising performances, in fact it shows a Mean Absolute Error of 3.62\% by using the feature set \textit{All\_features} and 3.67\% with \textit{Botometer+}'s features. 
%

We think that having an estimate of how many bots a human account follows, can be a first step to the fact-checking of what people reads on their dashboards and from whom they are reading. Additionally, such approaches can help researchers and social media administrators to (i) identify potential targets of misinformation campaigns in advance 
and (ii) contributing to increase the usefulness, credibility and effectiveness of social media.

Further efforts have to be devoted to this research topic; as future work it can be interesting to focus on the nature and quality of the information that humans with a large percentage of \textit{bot-followees} create and/or contribute to diffuse.

\section{Acknowledgements}
We would like to thank my supervisor Rocco De Nicola and  Catia Trubiani for the useful discussions and suggestions regarding this work.

\bibliographystyle{plain}
\newpage
\bibliography{0-easychair.bib}

\begin{thebibliography}{10}

\bibitem{aha1991instance}
David~W Aha, Dennis Kibler, and Marc~K Albert.
\newblock Instance-based learning algorithms.
\newblock {\em Machine learning}, 6(1):37--66, 1991.

\bibitem{amina2019bibliometric}
Shehu Amina, Raul Vera, Tooska Dargahi, and Ali Dehghantanha.
\newblock A bibliometric analysis of botnet detection techniques.
\newblock In {\em Handbook of Big Data and IoT Security}, pages 345--365.
  Springer, 2019.

\bibitem{aneez2019reuters}
Zeenab Aneez, Taberez~Ahmed Neyazi, Antonis Kalogeropoulos, and Rasmus~Kleis
  Nielsen.
\newblock Reuters institute india digital news report.
\newblock {\em Reuters Institute for the Study of Journalism/India Digital News
  Report. Retrieved from https://reutersinstitute. politics. ox. ac.
  uk/sites/default/files/2019-03/India\_DNR\_FINAL. pdf on}, 26:19, 2019.

\bibitem{Atkeson1996}
C.~Atkeson, A.~Moore, and S.~Schaal.
\newblock Locally weighted learning.
\newblock {\em AI Review}, 1996.

\bibitem{avvenuti2017}
Marco Avvenuti, Salvatore Bellomo, Stefano Cresci, {Mariantonietta Noemi} {La
  Polla}, and Maurizio Tesconi.
\newblock Hybrid crowdsensing: {A} novel paradigm to combine the strengths of
  opportunistic and participatory crowdsensing.
\newblock In {\em Proceedings of the 26th International Conference on World
  Wide Web Companion, Perth, Australia, April 3-7, 2017}, pages 1413--1421,
  2017.

\bibitem{balestrucci2019identification}
Alessandro Balestrucci, Rocco De~Nicola, Omar Inverso, and Catia Trubiani.
\newblock Identification of credulous users on twitter.
\newblock In {\em Proceedings of the 34th ACM/SIGAPP Symposium on Applied
  Computing}, pages 2096--2103. ACM, 2019.

\bibitem{balestrucci2019you}
Alessandro Balestrucci, Rocco De~Nicola, Marinella Petrocchi, and Catia
  Trubiani.
\newblock Do you really follow them? automatic detection of credulous twitter
  users.
\newblock {\em arXiv preprint arXiv:1909.03851 - to appear in Proceedings of
  IDEAL 2019}, 2019.

\bibitem{bessi2016social}
Alessandro Bessi and Emilio Ferrara.
\newblock Social bots distort the 2016 us presidential election online
  discussion.
\newblock {\em First Monday}, 21(11-7), 2016.

\bibitem{bradshaw2017troops}
Samantha Bradshaw and Philip Howard.
\newblock Troops, trolls and troublemakers: A global inventory of organized
  social media manipulation.
\newblock {\em Oxford Internet Institute}, 2017.

\bibitem{breiman2001random}
Leo Breiman.
\newblock Random forests.
\newblock {\em Machine learning}, 45(1):5--32, 2001.

\bibitem{costera2015checking}
Irene Costera~Meijer and Tim Groot~Kormelink.
\newblock Checking, sharing, clicking and linking: Changing patterns of news
  use between 2004 and 2014.
\newblock {\em Digital Journalism}, 3(5):664--679, 2015.

\bibitem{Cresci15fame}
Stefano Cresci, Roberto {Di Pietro}, Marinella Petrocchi, Angelo Spognardi, and
  Maurizio Tesconi.
\newblock Fame for sale: Efficient detection of fake {T}witter followers.
\newblock {\em Decision Support Systems}, 80:56--71, 2015.

\bibitem{cresci17paradigm}
Stefano Cresci, Roberto {Di Pietro}, Marinella Petrocchi, Angelo Spognardi, and
  Maurizio Tesconi.
\newblock The paradigm-shift of social spambots: Evidence, theories, and tools
  for the arms race.
\newblock In {\em Proceedings of the 26th International Conference on World
  Wide Web Companion, Perth, Australia, April 3-7, 2017}, pages 963--972, 2017.

\bibitem{daniel2019bots}
Florian Daniel, Cinzia Cappiello, and Boualem Benatallah.
\newblock Bots acting like humans: understanding and preventing harm.
\newblock {\em IEEE Internet Computing}, 23(2):40--49, 2019.

\bibitem{davis2016botornot}
Clayton~Allen Davis, Onur Varol, Emilio Ferrara, Alessandro Flammini, and
  Filippo Menczer.
\newblock Botornot: A system to evaluate social bots.
\newblock In {\em Proceedings of the 25th International Conference Companion on
  World Wide Web}, pages 273--274. International World Wide Web Conferences
  Steering Committee, 2016.

\bibitem{efthimion2018supervised}
Phillip~George Efthimion, Scott Payne, and Nicholas Proferes.
\newblock Supervised machine learning bot detection techniques to identify
  social twitter bots.
\newblock {\em SMU Data Science Review}, 1(2):5, 2018.

\bibitem{ferrara2017disinformation}
Emilio Ferrara.
\newblock Disinformation and social bot operations in the run up to the 2017
  french presidential election.
\newblock {\em First Monday}, 22(8), 2017.

\bibitem{ferrara2016rise}
Emilio Ferrara, Onur Varol, Clayton Davis, Filippo Menczer, and Alessandro
  Flammini.
\newblock The rise of social bots.
\newblock {\em Communications of the ACM}, 59(7):96--104, 2016.

\bibitem{frank2009conditional}
Eibe Frank and Remco~R Bouckaert.
\newblock Conditional density estimation with class probability estimators.
\newblock In {\em Asian Conference on Machine Learning}, pages 65--81.
  Springer, 2009.

\bibitem{Friedman1999}
J.H. Friedman.
\newblock Stochastic gradient boosting.
\newblock Technical report, Stanford University, 1999.

\bibitem{hall2009weka}
Mark Hall, Eibe Frank, Geoffrey Holmes, Bernhard Pfahringer, Peter Reutemann,
  and Ian~H Witten.
\newblock The weka data mining software: an update.
\newblock {\em ACM SIGKDD explorations newsletter}, 11(1):10--18, 2009.

\bibitem{howard2016bots}
Philip~N Howard and Bence Kollanyi.
\newblock Bots,\# strongerin, and\# brexit: computational propaganda during the
  uk-eu referendum.
\newblock {\em Available at SSRN 2798311}, 2016.

\bibitem{hsu2007paired}
Henry Hsu and Peter~A Lachenbruch.
\newblock Paired t test.
\newblock {\em Wiley encyclopedia of clinical trials}, pages 1--3, 2007.

\bibitem{iba1992induction}
Wayne Iba and Pat Langley.
\newblock Induction of one-level decision trees.
\newblock In {\em Machine Learning Proceedings 1992}, pages 233--240. Elsevier,
  1992.

\bibitem{keller2019social}
Tobias~R Keller and Ulrike Klinger.
\newblock Social bots in election campaigns: Theoretical, empirical, and
  methodological implications.
\newblock {\em Political Communication}, 36(1):171--189, 2019.

\bibitem{kempe2003maximizing}
David Kempe, Jon Kleinberg, and {\'E}va Tardos.
\newblock Maximizing the spread of influence through a social network.
\newblock In {\em Proceedings of the ninth ACM SIGKDD international conference
  on Knowledge discovery and data mining}, pages 137--146. ACM, 2003.

\bibitem{lazer2018science}
David~MJ Lazer, Matthew~A Baum, Yochai Benkler, Adam~J Berinsky, Kelly~M
  Greenhill, Filippo Menczer, Miriam~J Metzger, Brendan Nyhan, Gordon
  Pennycook, David Rothschild, et~al.
\newblock The science of fake news.
\newblock {\em Science}, 359(6380):1094--1096, 2018.

\bibitem{le1992ridge}
Saskia Le~Cessie and Johannes~C Van~Houwelingen.
\newblock Ridge estimators in logistic regression.
\newblock {\em Journal of the Royal Statistical Society: Series C (Applied
  Statistics)}, 41(1):191--201, 1992.

\bibitem{li2014hiding}
S~Li and ATS Ho.
\newblock Hiding information in a digital environment.
\newblock {\em Publication No. WO/2016/075459, International Application No.
  PCT/GB2015/053412, GB application filed on 11th November}, 2014.

\bibitem{Mackay1998}
David~J.C. Mackay.
\newblock Introduction to gaussian processes, 1998.

\bibitem{mitter2014categorization}
Silvia Mitter, Claudia Wagner, and Markus Strohmaier.
\newblock A categorization scheme for socialbot attacks in online social
  networks.
\newblock {\em arXiv preprint arXiv:1402.6288}, 2014.

\bibitem{moolayil2019learn}
Jojo Moolayil, Moolayil, and Suresh John.
\newblock {\em Learn Keras for Deep Neural Networks}.
\newblock Springer, 2019.

\bibitem{mushtaq2019distributed}
Atif Mushtaq, Todd Rosenberry, Ashar Aziz, and Ali Islam.
\newblock Distributed systems and methods for automatically detecting unknown
  bots and botnets, February~5 2019.
\newblock US Patent 10,200,384.

\bibitem{newman2017reuters}
Nic Newman, Richard Fletcher, Antonis Kalogeropoulos, David Levy, and
  Rasmus~Kleis Nielsen.
\newblock Reuters institute digital news report 2017.
\newblock {\em Reuters Institute for the Study of Journalism}, 2017.

\bibitem{newman2016digital}
Nic Newman, Richard Fletcher, Antonis Kalogeropoulos, David~AL Levy, and
  Rasmus-Kleis Nielsen.
\newblock Digital news report 2016.
\newblock {\em Reuters Institute for the study of Journalism}, 2016.

\bibitem{pal1992multilayer}
Sankar~K Pal and Sushmita Mitra.
\newblock Multilayer perceptron, fuzzy sets, and classification.
\newblock {\em IEEE Transactions on neural networks}, 3(5):683--697, 1992.

\bibitem{quinlan1987simplifying}
J.~Ross Quinlan.
\newblock Simplifying decision trees.
\newblock {\em International Journal of Human-Computer Studies},
  27(3):221--234, 1987.

\bibitem{Quinlan1992}
Ross~J. Quinlan.
\newblock Learning with continuous classes.
\newblock In {\em 5th Australian Joint Conference on Artificial Intelligence},
  pages 343--348, Singapore, 1992. World Scientific.

\bibitem{sandy2017can}
Christopher Sandy, Patrice Rusconi, and Shujun Li.
\newblock Can humans detect the authenticity of social media accounts? on the
  impact of verbal and non-verbal cues on credibility judgements of twitter
  profiles.
\newblock In {\em 2017 3rd IEEE International Conference on Cybernetics
  (CYBCONF)}, pages 1--8. IEEE, 2017.

\bibitem{savage2016botivist}
Saiph Savage, Andres Monroy-Hernandez, and Tobias H{\"o}llerer.
\newblock Botivist: Calling volunteers to action using online bots.
\newblock In {\em CSCW}. ACM, 2016.

\bibitem{scuse2007weka}
David Scuse and Peter Reutemann.
\newblock Weka experimenter tutorial for version 3-5-5.
\newblock {\em University of Waikato}, 2007.

\bibitem{DBLP:conf/websci/ShenCGLYL19}
Tracy~Jia Shen, Robert Cowell, Aditi Gupta, Thai Le, Amulya Yadav, and Dongwon
  Lee.
\newblock How gullible are you?: Predicting susceptibility to fake news.
\newblock In {\em Proceedings of the 11th {ACM} Conference on Web Science,
  WebSci 2019, Boston, MA, USA, June 30 - July 03, 2019}, pages 287--288, 2019.

\bibitem{Shevade1999}
S.K. Shevade, S.S. Keerthi, C.~Bhattacharyya, and K.R.K. Murthy.
\newblock Improvements to the smo algorithm for svm regression.
\newblock In {\em IEEE Transactions on Neural Networks}, 1999.

\bibitem{suarez2016influence}
Pablo Su{\'a}rez-Serrato, Margaret~E Roberts, Clayton Davis, and Filippo
  Menczer.
\newblock On the influence of social bots in online protests.
\newblock In {\em International Conference on Social Informatics}, pages
  269--278. Springer, 2016.

\bibitem{Varol17}
Onur Varol, Emilio Ferrara, {Clayton A.} Davis, Filippo Menczer, and Alessandro
  Flammini.
\newblock Online human-bot interactions: Detection, estimation, and
  characterization.
\newblock In {\em Proceedings of the Eleventh International Conference on Web
  and Social Media, {ICWSM} 2017, Montr{\'{e}}al, Qu{\'{e}}bec, Canada, May
  15-18, 2017.}, pages 280--289, 2017.

\bibitem{wagner2012social}
Claudia Wagner, Silvia Mitter, Christian K{\"o}rner, and Markus Strohmaier.
\newblock When social bots attack: Modeling susceptibility of users in online
  social networks.
\newblock In {\em \# MSM}, pages 41--48, 2012.

\bibitem{wald2013predicting}
Randall Wald, Taghi~M Khoshgoftaar, Amri Napolitano, and Chris Sumner.
\newblock Predicting susceptibility to social bots on twitter.
\newblock In {\em 2013 IEEE 14th International Conference on Information Reuse
  \& Integration (IRI)}, pages 6--13. IEEE, 2013.

\bibitem{willmott2005advantages}
Cort~J Willmott and Kenji Matsuura.
\newblock Advantages of the mean absolute error (mae) over the root mean square
  error (rmse) in assessing average model performance.
\newblock {\em Climate research}, 30(1):79--82, 2005.

\bibitem{yang2019arming}
Kai-Cheng Yang, Onur Varol, Clayton~A Davis, Emilio Ferrara, Alessandro
  Flammini, and Filippo Menczer.
\newblock Arming the public with artificial intelligence to counter social
  bots.
\newblock {\em Human Behavior and Emerging Technologies}, 1(1):48--61, 2019.

\bibitem{zhou2018fake}
Xinyi Zhou and Reza Zafarani.
\newblock Fake news: A survey of research, detection methods, and
  opportunities.
\newblock {\em arXiv preprint arXiv:1812.00315}, 2018.

\end{thebibliography}
\appendix
\section{Appendix}

\subsection{Feature sets}
\label{app:feat}
\begin{table}[htbp]
	\begin{center}
		\begin{tabular}{lcl}
			\hline
{\textbf{Name}} & \multicolumn{1}{c}{\textbf{Original Features}} & \multicolumn{1}{c}{\textbf{\makecell{Features alteration}}} \\ \hline
			\makecell[l]{\textbf{Botometer+}\\used in\cite{balestrucci2019you}\\ From Botometer\cite{Varol17}}& \makecell{Macro categories scores:\\
			Sentiment, Friend, \\
			User, Content, \\
			Temporal, Net } & \makecell[l]{ (+) CAP (ENG and UNI)\\
			(+) Score (ENG and UNI)\\
			(+) \#Tweets4WS 
			\\
			(+) \#Mentions4WS 
			}\\ \hline
			\makecell[l]{\textbf{ClassA-}\\used in\cite{balestrucci2019you}\\ From Fake Follower\cite{Cresci15fame}\\(Class A)}&  \makecell{friends/(followersˆ2), age, \\
			\#tweets, profile has name, \\
			has URL in profile, following rate, \\
			default image after 2 months, \\
			belongs to a list, profile has image, \\
			friends/followers$\geq$50, \\`bot' in biography, \#friends,\\
			2$\times$ followers $\geq$ friends, \\
			\#followers, friends/followers$\simeq$100, \\
			no bio \& no location \& friends$\geq$100, \\
			has address, has biography }& \makecell[l]{(-) duplicated pictures \\  of the profile }\\ \hline
		\end{tabular}
	\end{center}
	\caption{Feature sets}
	\label{tab:features}
\end{table}

\end{document}